\documentclass[12pt]{iopart}

\usepackage{graphicx}

\begin{document}

\title[Dimitrijevi\'c \etal]{Continuous reversal of Hanle resonances of counter-propagating pulse and continuous-wave field}

\author{Jelena Dimitrijevi\'c, Du\v san Arsenovi\'c and Branislav M Jelenkovi\'c}
\address{Institute of Physics, University of Belgrade, Pregrevica 118, 10080 Belgrade, Serbia}
\ead{jelena.dimitrijevic@ipb.ac.rs}

\begin{abstract}

In this work we study propagation dynamics of the two counter-propagating lasers, the continuous-wave (CW) laser and the pulse of another laser, when both lasers are tuned to the $F_{g}=2 \rightarrow F_{e}=1$ transition in $^{87}$Rb, and therefore can develop Hanle electromagnetically induced transparency (EIT) in Rb vapor. We calculate transmission of both lasers as a function of applied magnetic field, and investigate how the propagation of the pulse affects the transmission of the CW laser. And vice versa, we have found conditions when the Gaussian pulse can either pass unchanged, or be significantly absorbed in the vacuum Rb cell. This configuration is therefore suitable for the convenient control of the pulse propagation and the system is of interest for optically switching of the laser pulses. In terms of the corresponding shapes of the coherent Hanle resonances, this is equivalent to turning the coherent resonance from Hanle EIT into electromagnetically induced  absorption (EIA) peak. There is the 
range of intensities of both CW laser and the laser pulse when strong drives of atomic coherences allow two lasers to interact with each other through atomic coherence and can simultaneously reverse signs of Hanle resonances of both.

\end{abstract}


\pacs{42.50.Gy,42.50.Nn}
\submitto{Laser Physics}
\maketitle

\section{Introduction}

Interaction of atoms with lasers has been one of the most studied subjects during the last decades. Coherent phenomena in atoms like the coherent population trapping \cite{aspect88}, Hanle effect \cite{moruzzi91} and related phenomena, electromagnetically induced transparency (EIT) \cite{arimondo96} and electromagnetically induced absorption (EIA) \cite{akulshin98}, have been widely studied under various conditions. EIT and EIA can be induced in different atomic schemes. It can be pump-probe configuration where two lasers couple two or more hyperfine (or Zeeman) levels. The other way to probe the distribution of atomic coherences is to apply single optical field, while the transmission of the field is measured as a function of the magnetic field that varies energy of Zeeman sublevels, so called Hanle configuration.

EIT and EIA narrow resonances and steep dispersion in the narrow spectral bandwidth of their resonances represent the unique properties of atomic systems. The ability to switch from EIT to EIA could provide a new technique to manipulate over the properties of a medium. Yu \etal \cite{yu10} demonstrated transformation from the EIT to the EIA in the Hanle configuration when the polarization of a traveling wave changed gradually from linear to circular. This sign reversal was connected with the weak residual transverse magnetic field perpendicular to laser propagation. Bae \etal \cite{bae11} recently presented the continuous control of the light group velocity from subluminal to superluminal propagation by using the standing-wave coupling field in the transition of the $\Lambda$-type system of $^{87}$Rb atoms.  When the coupling field changed from a traveling wave to a standing wave by changing the power of counter-propagating laser field, and the speed of the probe pulse changed from 
subluminal 
to superluminal  propagation.

The counter-propagating geometry and the resulting variations of transmission and refractive index have also been explored. Chanu  \etal \cite{chanu12} recently showed that it is possible to reverse sign of subnatural resonances in a (degenerate) three-level Λ system. They observed the D2 line of Rb, in a room temperature vapor cell, and change is obtained by turning on a second control beam counter-propagating with respect to the first beam.  The role of the different possible subsystems created in this configuration was analyzed \cite{pandey13}, together with the possibility of tuning the strength of individual subsystems by changing the polarization of the control lasers.

Counter-propagation dynamics for the Hanle configuration has also been investigated  \cite{zibrov07, zhukov09, brazhnikov10, brazhnikov11}. Brazhnikov \etal \cite{brazhnikov10, brazhnikov11} recently analyzed counter-propagating geometry of the two CW laser fields in the Hanle configuration. They used polarization method to reverse the sign of Hanle resonance of the CW field, and numerical and analytical calculations were performed for the simple three-level schemes \cite{brazhnikov10, brazhnikov11}.

In this paper, we present development of Hanle resonances of two counter-propagating lasers, the CW and of the Gaussian pulsed laser. Lasers, having orthogonal linear polarizations, both couple the $F_{g}=2 \rightarrow F_{e}=1$ transition in $^{87}$Rb D1 line. Analysis is performed by numerical solving the set of Maxwell-Bloch equations for the same transition including all the magnetic sublevels and for the cold atoms. We have found range of intensities for both lasers such that they can influence transmission of each other, that is, they can continuously reverse the sign of their resonances from EIT to EIA and \textit{vice versa}, while the pulse is passing through the Rb cell. We are interested for dynamics in two special cases:  when the pulse of one laser reverses the sign of  the second, CW laser, and when both lasers can simultaneously switch each others' resonances signs.

\section{\label{model} Theoretical model}

\begin{figure}[h]
\centering\includegraphics[width=0.5 \textwidth]{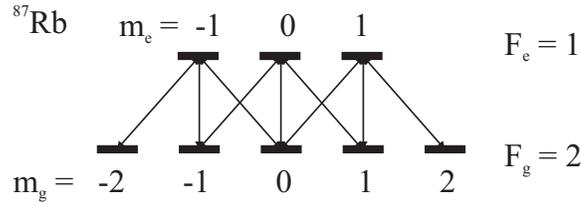}
\caption{\label{sema}
$F_{g}=2$ and $F_{e}=1$ hyperfine levels with the notation of magnetic sublevels.}
\end{figure}

Here we describe the model that calculates the transmissions of two lasers which act simultaneously on Rb atoms. Namely, there is constant interaction with one laser, CW, and, for the some time, also with the pulse of second laser.  The set of Maxwell-Bloch equations (MBEs) is solved for all magnetic sublevels of the $F_{g} =2 \rightarrow F_{e} = 1$ transition (see figure \ref{sema}). From the optical Bloch equations (OBEs):
\begin{equation}
\label{OBE}
\frac{d \hat{\rho}(t)}{dt}
=-\frac{i}{\hbar}[\hat{H}_{0},\hat{\rho}(t)]-\frac{i}{\hbar}[\hat{H}_{I},\hat{\rho}(t)]-
\hat{SE}\hat{\rho}(t)-\gamma\hat{\rho}(t)+\gamma\hat{\rho}_{0}
\end{equation}
we calculate the evolution of density matrix $\hat{\rho}$.  The diagonal elements of the density matrix, $\rho_{g_i,g_i}$ and $\rho_{e_i,e_i}$ are the populations, $\rho_{g_i,g_j}$ and $\rho_{e_i,e_j}$ are the Zeeman coherences, and $\rho_{g_i,e_j}$ and $\rho_{e_i,g_j}$ are the optical coherences. Here, indices $g$ and $e$ are  for the ground and the excited levels, respectively.

MBEs are solved for the Hanle configuration i.e. when the external magnetic field $B$ is varied around the zero value, as described by the Hamiltonian part $\hat{H}_{0}$. The direction of $\vec{B}$ is the direction of laser propagation and it is also the quantization axis. The Zeeman splitting of the magnetic sublevels is $E_{g(e)}=\mu_{B}l_{F_{g(e)}}m_{g(e)}B$, where $m_{g(e)}$ are the magnetic quantum numbers of the ground and excited levels, $\mu_{B}$ is the Bohr magneton and $l_{F_{g,e}}$ is the Lande gyromagnetic factor for the  hyperfine levels. In equation \ref{OBE}, $\hat{SE}$ is the spontaneous emission operator, the rate of which is $\Gamma$, and in our calculations, we have taken into account that the transition $F_{g}=2 \rightarrow F_{e}=1$ is open. The term $\gamma\hat{\rho}$ describes the relaxation of all density matrix elements, due to a finite time that an atom spends in the laser beam. The continuous flux of atoms entering the laser beams is described  with the  term $\gamma\hat{\rho}_{0}$. We 
assume equal population of the ground Zeeman sublevels for these atoms. The role of the laser detuning (and Doppler broadening) is not discussed.

Atoms are interacting with the CW laser, and, during the limited time, they simultaneously also interact with another, pulsed laser. These interactions are described with Hamiltonian $\hat{H}_{I}$. The electric field vector represents the sum of two fields:
\begin{equation}
\label{Evec1}
\vec{E}(t,z)=\sum_{l} [E_x^{l}\cos( \omega^{l} t - k^{l} z+\varphi _x^{l}) \vec{e}_{x} + E_y^{l} \cos( \omega ^{l} t - k^{l} z+\varphi _y^{l}) \vec{e}_{y}].
\end{equation}
In equation \ref{Evec1}, $\omega^{l}>0$ are the laser's ($l$) angular frequencies $\omega^{l}=\pm c k^{l}$, $k^{l}$ are wave vectors, where we take $k^{l}<0$ when the propagation is towards the negative direction of the $z$-axis, and $c$ is the speed of light. $E_x^{l}, E_y^{l}$ are the real Descartes components of amplitudes of the electric field while $\varphi _x^{l}, \varphi _y^{l}$ are associated phases and also real quantities.

We introduce the following substitution:
\begin{eqnarray}
\label{rule2}
  E_{++}^{l} &=& \frac{-E_x^{l} e^{+i \varphi _x^{l}}+i E_y^{l} e^{+i \varphi _y^{l}}}{2\sqrt{2}},\\ \nonumber
  E_{+-}^{l} &=& \frac{-E_x^{l} e^{-i \varphi _x^{l}}+i E_y^{l} e^{-i \varphi _y^{l}}}{2\sqrt{2}},\\ \nonumber
  E_{-+}^{l} &=& \frac{E_x^{l} e^{+i \varphi _x^{l}}+i E_y^{l} e^{+i \varphi _y^{l}}}{2\sqrt{2}},\\ \nonumber   E_{--}^{l} &=& \frac{E_x^{l} e^{-i \varphi _x^{l}}+i E_y^{l} e^{-i \varphi _y^{l}}}{2\sqrt{2}}.
\end{eqnarray}
With this substitution, the electric field vector stands:
\begin{eqnarray}
\label{Evec6}
  &&\vec{E}(t,z) =\sum_{l}\vec{E^l}(t,z) \sum_{l}[
      e^{i(\omega^{l}t-k^{l}z)}  \vec{u}_{+1} E_{++}^{l} +
      e^{i(\omega^{l}t-k^{l}z)}  \vec{u}_{-1} E_{-+}^{l} \\ \nonumber
  &+& e^{-i(\omega^{l}t-k^{l}z)} \vec{u}_{+1} E_{+-}^{l}+
      e^{-i(\omega^{l}t-k^{l}z)} \vec{u}_{-1} E_{--}^{l}],
\end{eqnarray}
where $ \vec{u}_{+1}$,  $ \vec{u}_{-1}$ and $ \vec{u}_{0}$ are  spherical unit vectors \cite{edmonds96}.

For the case of the two counter-propagating lasers that couple the same transition, we apply the multi-mode Floquet theory \cite{chu04}. We use the approximation with the zeroth-order harmonics for the ground-state and excited-state density matrices, and up to the first-order harmonics for the optical coherences:
\begin{eqnarray}
\label{substitution1}
\rho_{g_{i},e_{j}} &=&\sum_{l} e^{i(\omega^{l}t-k^{l}z)} \tilde{\rho}_{g_{i},e_{j}}^{l},\\ \nonumber
\rho_{e_{i},g_{j}} &=&\sum_{l} e^{-i(\omega^{l}t-k^{l}z)} \tilde{\rho}_{e_{i},g_{j}}^{l},
\end{eqnarray}
where the sum is taken over lasers that couple states $g_{i}$ and $e_{j}$.

The macroscopic polarization of the atomic medium $\vec{P}(t,z)=N_{c} e Tr[\hat{\rho} \hat{\vec{r}}]$ is calculated as:
\begin{eqnarray}
\label{Pvec2}
   \vec{P}(t,z) &=& N_{c}\sum_{l}[
      e^{i(\omega^{l}t-k^{l}z)}  (\vec{u}_{+1} P_{++}^{l} +\vec{u}_{-1} P_{-+}^{l})\\ \nonumber
    &+&  e^{-i(\omega^{l}t-k^{l}z)} (\vec{u}_{+1} P_{+-}^{l}+ \vec{u}_{-1} P_{--}^{l})],
\end{eqnarray}
where we introduced new quantities:
\begin{eqnarray}
\label{Pvars}
  P_{++}^{l} &=&  \sum_{g_{i} \leftrightarrow e_{j}} \tilde{\rho}_{g_{i},e_{j}}^{l} \mu_{g_{i},e_{j},+1},\\ \nonumber
  P_{+-}^{l} &=&  \sum_{e_{i} \leftrightarrow g_{j}} \tilde{\rho}_{e_{l},g_{j}}^{l} \mu_{g_{i},e_{j},+1},\\ \nonumber
  P_{-+}^{l} &=&  \sum_{g_{i} \leftrightarrow e_{j}} \tilde{\rho}_{g_{i},e_{j}}^{l} \mu_{g_{i},e_{j},-1},\\ \nonumber
  P_{--}^{l} &=&  \sum_{e_{i} \leftrightarrow g_{j}} \tilde{\rho}_{e_{i},g_{j}}^{l}
  \mu_{g_{i},e_{j},-1}.
\end{eqnarray}
The sum is taken over the dipole-allowed transitions induced by lasers $(ls)$.

MBEs are solved for $E_{++}^{l}, E_{+-}^{l},  E_{-+}^{l}, E_{--}^{l}$ (given by equation \ref{rule2}) which are  complex amplitudes of the fields taking into account the relations $(E_{++}^{l})^{*}=-E_{--}^{l}, (E_{+-}^{l})^{*}=-E_{-+}^{l}$. MBEs, for the propagation along the positive direction of the $z$-axis, are:
\begin{eqnarray}
\label{MBpoz}
  (\frac{\partial}{\partial z} +\frac{1}{c}\frac{\partial}{\partial t}) E_{++}^{l} &=&
 -i\frac{k^{l}N_{c}}{2 \varepsilon_{0}} P_{++}^{l},\\ \nonumber
   (\frac{\partial}{\partial z} +\frac{1}{c}\frac{\partial}{\partial t}) E_{-+}^{l} &=&
 -i\frac{k^{l}N_{c}}{2 \varepsilon_{0}} P_{-+}^{l},\\ \nonumber
 \ (\frac{\partial}{\partial z} +\frac{1}{c}\frac{\partial}{\partial t}) E_{+-}^{l}&=&
 +i\frac{k^{l}N_{c}}{2 \varepsilon_{0}} P_{+-}^{l},\\ \nonumber
  \ (\frac{\partial}{\partial z} +\frac{1}{c}\frac{\partial}{\partial t}) E_{--}^{l}&=&
 +i\frac{k^{l}N_{c}}{2 \varepsilon_{0}} P_{--}^{l},
\end{eqnarray}
while for the propagation along the negative direction of the $z$-axis are:
\begin{eqnarray}
\label{MBneg}
  (-\frac{\partial}{\partial z} +\frac{1}{c}\frac{\partial}{\partial t}) E_{++}^{l} &=&
 -i\frac{k^{l}N_{c}}{2 \varepsilon_{0}} P_{++}^{l},\\ \nonumber
   (-\frac{\partial}{\partial z} +\frac{1}{c}\frac{\partial}{\partial t}) E_{-+}^{l} &=&
 -i\frac{k^{l}N_{c}}{2 \varepsilon_{0}} P_{-+}^{l},\\ \nonumber 
 \ (-\frac{\partial}{\partial z} +\frac{1}{c}\frac{\partial}{\partial t}) E_{+-}^{l}&=&
 i\frac{k^{l}N_{c}}{2 \varepsilon_{0}} P_{+-}^{l},\\ \nonumber
  \ (-\frac{\partial}{\partial z} +\frac{1}{c}\frac{\partial}{\partial t}) E_{--}^{l}&=&
 i\frac{k^{l}N_{c}}{2 \varepsilon_{0}} P_{--}^{l}.
\end{eqnarray}

In the following text, we present results for the transmission of lasers, that is for the averaged power of the laser (l) electromagnetic fields:
\begin{equation}
\label{intenzitet1}
    I^l=c \varepsilon_{0} \langle \vec{E^l} \cdot \vec{E^l} \rangle
\end{equation}
From equation \ref{Evec6} we have
\begin{eqnarray}
\label{intenzitet3}
  I^l &=& c \varepsilon_{0}\langle  \{\vec{u}_{+1} [E_{++}^{l}e^{i(\omega^{l}t-k^{l}z)}+E_{+-}^{l} e^{-i(\omega^{l}t-k^{l}z)}]\\ \nonumber
  &+&\vec{u}_{-1} [E_{-+}^{l} e^{i(\omega^{l}t-k^{l}z)} +E_{--}^{l} e^{-i(\omega^{l}t-k^{l}z)}]\}^{2}\rangle \\ \nonumber
  &=&-2 c \varepsilon_{0}\  \{ E_{++}^{l} E_{+-}^{l}+ E_{-+}^{l} E_{--}^{l} \}.
\end{eqnarray}

\section{Results and discussion}

In this section we present effects on the Hanle EIT of the CW laser when the counter-propagating laser pulse, tuned to the same $F_{g} =2 \rightarrow F_{e} = 1$ transition, passes the Rb cell, overlapping the CW laser. And reversely, we present results when the CW laser controls the transmission and the Hanle EIT of the laser pulse.

\begin{figure}[htbp]
\centering\includegraphics[width=0.5 \textwidth]{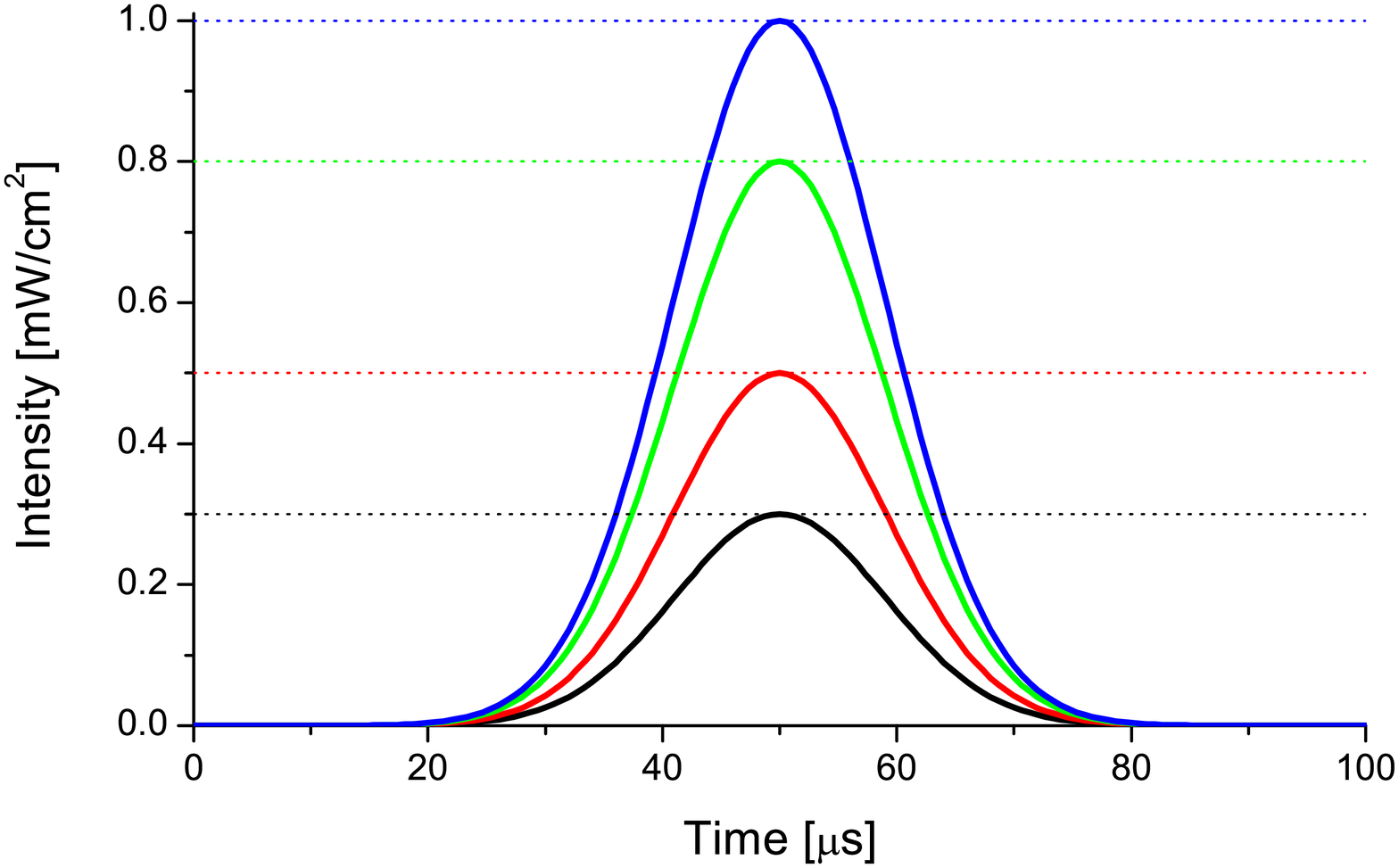}
\caption{\label{slikaP&CW}
Figure shows the temporal waveforms of the laser amplitudes that we used in our calculations. The straight lines indicate the amplitudes of the CW laser $I_{CW}^{0}$ and full-lines temporal wave-forms of pulses $I_{pulse}$.}
\end{figure}

Transmissions of lasers are calculated for the values of the external magnetic field near zero, i.e. around the EIT resonance. Polarizations of both lasers are linear and orthogonal. We take atoms concentration in the cell $N_{c}=10^{14}$ m$^{-3}$, length of the cell is $10$ cm, $\gamma=0.001\ \Gamma$ and the spontaneous emission rate is $\Gamma=2\pi\ 5.75$ MHz. The temporal shape of the laser pulse is Gaussian $I_{pulse}^0 e^{-\frac{(t-t_0)^2}{\sigma^2}}$ (see figure \ref{slikaP&CW}), where $\sigma=10 \mu s/ \sqrt{2 \ln 2}$ and $I_{pulse}^{0}$ is the intensity of the laser pulse, at the peak of the amplitude at $t_{0}=50$ $\mu$s.

Since both lasers are resonant with the $F_{g} =2 \rightarrow F_{e} = 1$ transition in $^{87}$Rb, each can independently induce EIT in the atomic vapor. As the pulse enters the cell, it's intensity increases, reaches it's maximal value $I_{pulse}^{0}$ and then decreases. At the mere beginning of the pulse, atoms are interacting with the CW laser only, and the transmission of the CW field shows Hanle EIT. As the intensity of pulse increases, atoms begin to interact with both superimposed fields. Resultant polarization, which is the sum of two orthogonal linear polarizations, can yield very different shapes of transmission resonances of both lasers.

As we see below, complete switching of the sign of transmission resonance, from transmission gain to absorption gain, can happen depending on the ratio of the pulse's peak intensity $I_{pulse}^{0}$ and the intensity of the CW field $I_{CW}^{0}$. We will next discuss two possible cases, $I_{pulse}^{0} < I_{CW}^{0}$ and $I_{pulse}^{0} > I_{CW}^{0}$.

\subsection{Hanle EIT resonances of the CW laser field in the presence of the weak laser pulse}

\begin{figure}[htbp]
\centering\includegraphics[width=0.8 \textwidth]{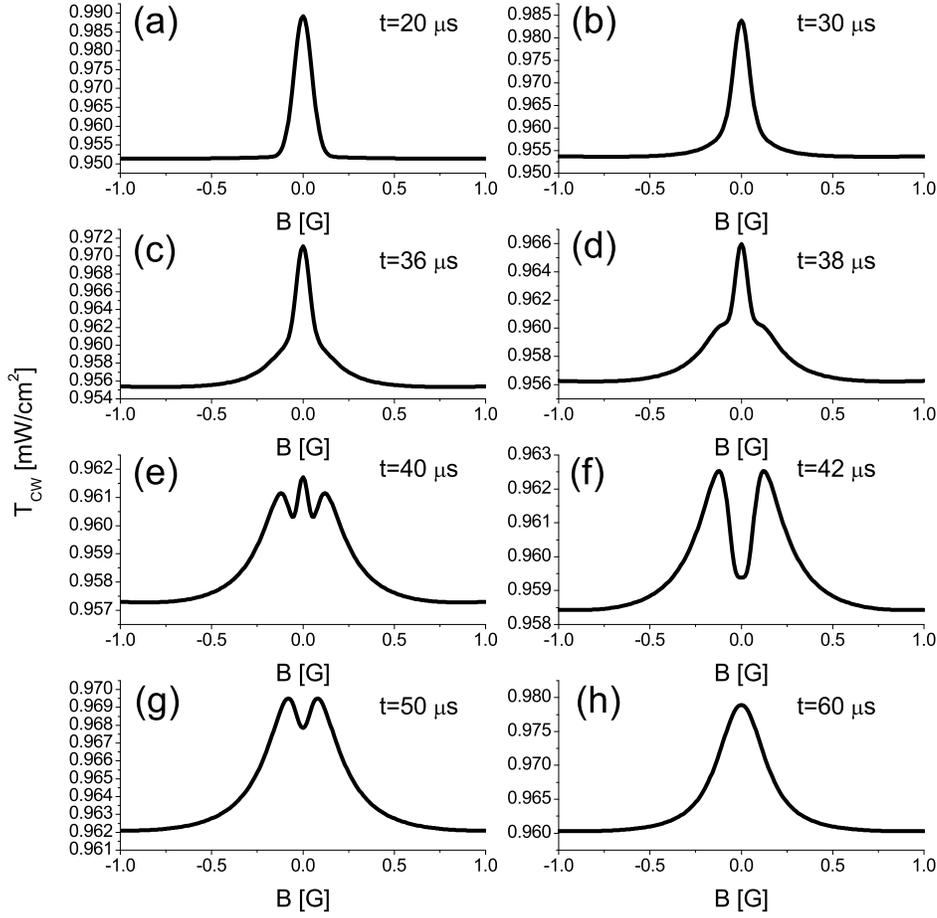}
\caption{\label{VCW}
Transmission of the CW laser as a function of the magnetic field $B$ at different times regarding the position of the pulse of the second laser in respect to the Rb cell: $t=20$ $\mu$s (a),  $t=30$ $\mu$s (b),  $t=36$ $\mu$s (c),  $t=38$ $\mu$s (d),  $t=40$ $\mu$s (e),  $t=42$ $\mu$s (f),   $t=50$ $\mu$s (g) and  $t=60$ $\mu$s (h). The CW laser's intensity is $I_{CW}^{0}=1$ mW/cm$^{2}$ and the intensity of the pulse at the maximum is $I_{pulse}^{0}=0.3$ mW/cm$^{2}$ (solid black and dashed blue curves in figure \ref{slikaP&CW}). Note that the first curve is before the pulse even enters the cell, while the curve (g) is at the time when the pulse is at the highest intensity, as can be seen from the figure \ref{slikaP&CW}.}
\end{figure}

In figure \ref{VCW} we show changes of the CW laser Hanle EIT as the counter-propagating and spatially overlapping laser pulse changes its intensity in the Rb cell. Results are for laser intensities, $I_{CW}^{0}=1$ mW/cm$^{2}$ and $I_{pulse}^{0}=0.3$ mW/cm$^{2}$ (solid black and dashed blue lines in figure \ref{slikaP&CW}).  As the laser pulse enters the cell the transmission of the CW laser still shows the Hanle EIT. Such is the case with the front of the pulse in the cell, that is for results presented in figures \ref{VCW} (a)-(c). As the pulse intensity increases, this EIT widens and structure begins to form at the resonance center. During the transient period, when the pulse intensity is increasing, there are instances when the CW transmission at the zero value of the external magnetic field, totally switches its transmission behavior. Instead of the maximum transmission, when the laser pulse is off, it is maximally absorbed in the cell. It changes the sign from EIT to EIA. With the back side of the pulse 
left in the cell, at the time $t=60$ $\mu$s, the transmission of the CW laser shows again EIT, and the resonance stays like this until  there is no pulse laser in the cell. 

For the choice of laser intensities considered here, $I \leq 1$ mW/cm$^{2}$ and with the condition $I_{pulse}^{0} \leq I_{CW}^{0}$, the transmission of the pulse laser is showing EIT at all times. Besides intensity broadening, the transmission of the pulse, as a function of the magnetic field, does not change significantly as the pulse passes. Results for the pulse's sign reversal are presented in the next subsection.


\begin{figure}[htbp]
\centering\includegraphics[width=0.8 \textwidth]{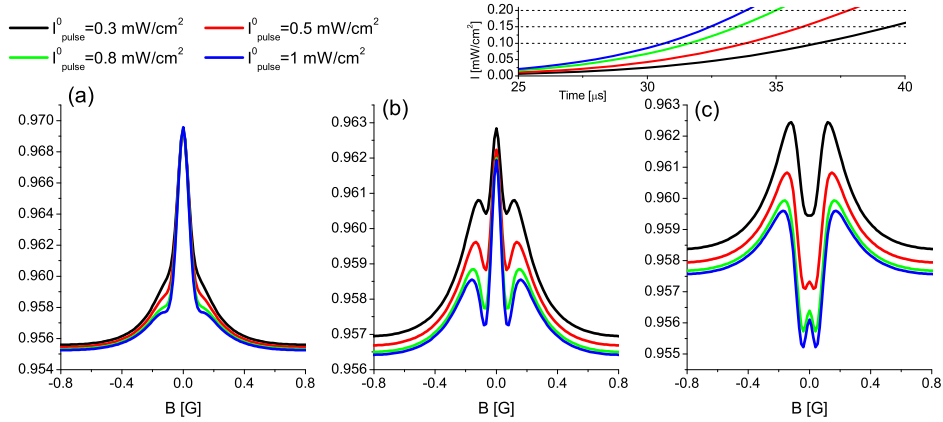}
\caption{\label{II}
Transmissions of the CW laser for times when the pulse is entering the cell with $3$ different intensities:  $I_{pulse}=0.1\ $ mW/cm$^{2}$ (a), $I_{pulse}=0.15\ $ mW/cm$^{2}$ (b) and $I_{pulse}=0.2\ $ mW/cm$^{2}$ (c). For each $I_{pulse}$ we present results for $4$ different rise time of the intensity, i.e., when the the pulse has maximum intensities: $I_{pulse}^{0}=0.3$ mW/cm$^{2}$ (black curves), $I_{pulse}^{0}=0.5$ mW/cm$^{2}$ (red curves),  $I_{pulse}^{0}=0.8$ mW/cm$^{2}$ (green curves) and $I_{pulse}^{0}=1$ mW/cm$^{2}$ (blue curves). The intensity of the CW laser is $I_{CW}^{0}=1$ mW/cm$^{2}$. The inset graph shows these $3$ chosen intensities from $4$ pulses as horizontal black dashed lines.}
\end{figure}

In figure \ref{II} we compare the transmissions of the CW laser when four pulses, with different maximum values, have the same intensity in the cell. That is, when pulses whose maximums are  $I_{pulse}^{0}=0.3$ mW/cm$^{2}$ (black curves), $I_{pulse}^{0}=0.5$ mW/cm$^{2}$ (red curves),  $I_{pulse}^{0}=0.8$ mW/cm$^{2}$ (green curves) and $I_{pulse}^{0}=1$ mW/cm$^{2}$ (blue curves) (see figure \ref{slikaP&CW}), have the same values of $I_{pulse}=0.1 $ mW/cm$^{2}$ (a), $I_{pulse}=0.15 $ mW/cm$^{2}$ (b) and $I_{pulse}=0.2 $ mW/cm$^{2}$ (c).

Results in figure \ref{II} show that besides the intensity, the slope of the rising front of the pulse also determines the transmissions of the CW laser. Though atoms are affected by the same pulse's intensity, as given in separate figures, the transmission of the CW field does not show identical results when curves are compared. From figure \ref{II} we see that evolution of the EIT/EIA passes through the similar stages as the pulse raises, but the speed of the evolution of the EIT/EIA depends on the pulse's slope  i.e. it's first derivative. This also means that sign reversal of the CW laser, for different Gaussian pulse (different maximal intensities), does not happen simultaneously.


\begin{figure}[htbp]
\centering\includegraphics[width=0.8 \textwidth]{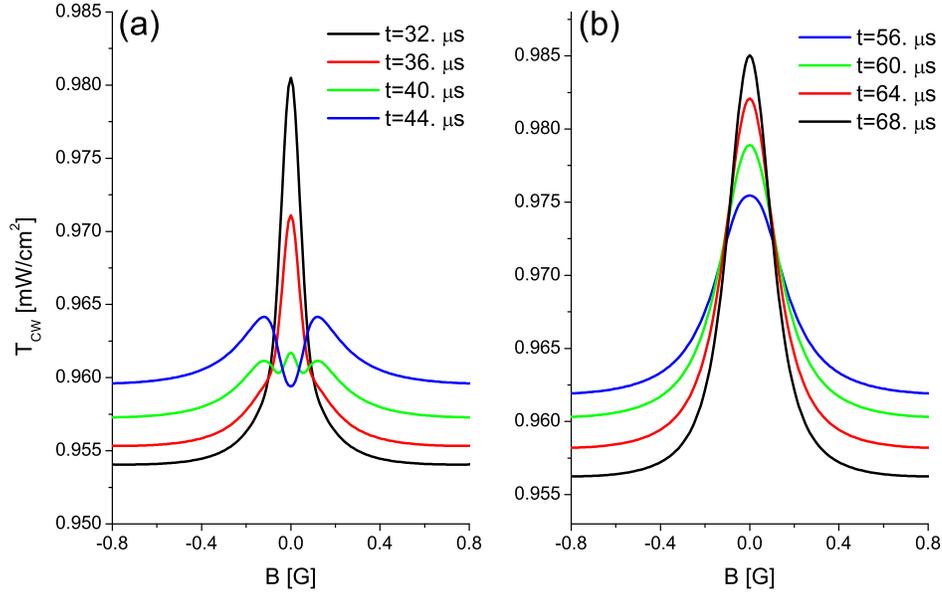}
\caption{\label{II0.3}
The transmissions of the CW laser, when the pulse laser takes $4$ intensities from the rising (a) and falling (b) edge of the pulse. Certain intensity is chosen from the symmetrical moments of the pulse: $t=32$ $\mu$s and $t=68$ $\mu$s (black curves),  $t=36$ $\mu$s and $t=64$ $\mu$s (red curves),  $t=40$ $\mu$s and $t=60$ $\mu$s (green curves) and $t=44$ $\mu$s and $t=56$ $\mu$s (blue curves). Intensity of the CW laser is  $I_{CW}^{0}=1$ mW/cm$^{2}$ and of the pulse laser is $I_{pulse}^{0}=0.3$ mW/cm$^{2}$ (solid black and dashed blue curves in figure \ref{slikaP&CW}). }
\end{figure}

In figure \ref{II0.3} we show that the rising and falling edges of the laser pulses, with the same intensity, have different effects on the CW laser transmission.  For the CW laser intensity of $I_{CW}^{0}=1$ mW/cm$^{2}$ and for the pulse laser maximum intensity of $I_{pulse}^{0}=0.3$ mW/cm$^{2}$, we present the CW laser transmission for four pairs of equal pulse laser intensity, each pair has the same intensity on two sides of the laser pulse. We choose the following pairs of intensities of the laser pulse, i.e., for the pulse instants: $t=32$ $\mu$s and $t=68$ $\mu$s (black curves),  $t=36$ $\mu$s and $t=64$ $\mu$s (red curves),  $t=40$ $\mu$s and $t=60$ $\mu$s (green curves) and $t=44$ $\mu$s and $t=56$ $\mu$s (blue curves).

Results given in figure \ref{II0.3} indicate that the transmissions of the CW field are not symmetrical in respect to the pulse maximum intensity. Depending whether CW laser is overlapped with the rising (figure \ref{II0.3} (a)) or the falling (figure \ref{II0.3} (b)) edge of the pulse, profiles have different waveforms, although the intensities are the same. Similar result of this "memory effect" has also been discussed by Ignesti \etal \cite{ignesti11}. It was theoretically predicted that spectral enlargement or compression process occurs when the probe pulse is overlapped by a coupling pulse field with a positive or negative temporal slope and experimental confirmation was obtained for a scheme in sodium atomic vapour \cite{ignesti11}. At the rising edge of the pulse, while the dark-state is being formed, there are rapid changes in the transmission of the CW laser. The profiles at the falling edge of the pulse are due to long-lived ground-state coherences, created at the earlier time of the pulse 
propagation.  
After certain time the dark-state is established and further decrease of the pulse's intensity does not affect the formed dark-state.

\subsection{Switching signs of both lasers transmissions}

Our analysis indicate that the simultaneous switching of signs of transmission resonances of both CW and the pulse laser fields is when the intensity of the CW field is comparable to the pulse laser intensity, that is $I_{CW}^{0} \leq I_{pulse}^{0}$. For this ratio of intensities, when the pulse laser intensity after certain time reaches the intensity of the CW laser, the intensities of two lasers are comparable (see figure \ref{slikaP&CW}) and reverse of signs of both are possible.

\begin{figure}[htbp]
\centering\includegraphics[width=0.8 \textwidth]{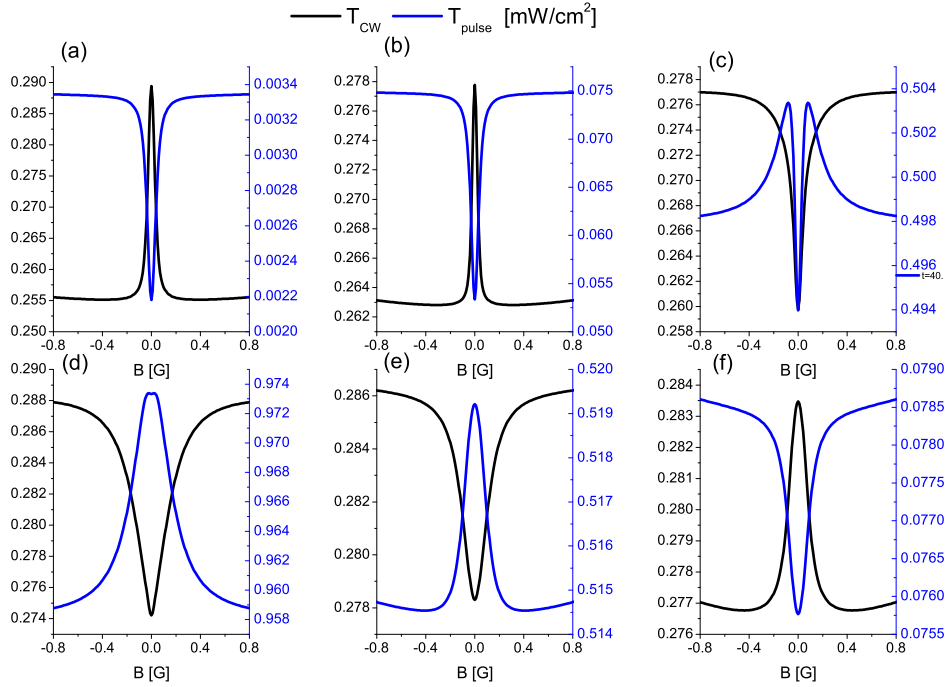}
\caption{\label{Voba}
Transmissions of both CW laser and the pulse as a function of the magnetic field: CW laser (black curves and black $y$-axis) and pulse laser (blue curves and blue y-axis). The amplitude of the CW laser's electric field is $I_{CW}^{0}=0.3$ mW/cm$^{2}$ and the pulse's maximum values is $I_{pulse}^{0}=1$ mW/cm$^{2}$  (solid blue and dashed black curves in figure \ref{slikaP&CW}). Results are presented for $6$  values of the intensity of the Gaussian laser pulse as it passes through the cell, from $t=20$ $\mu$s to $t=70$ $\mu$s in steps of $10$ $\mu$s: $t=20$ $\mu$s (a), $t=30$ $\mu$s (b), $t=40$ $\mu$s (c), $t=50$ $\mu$s (d), $t=60$ $\mu$s (e) and $t=70$ $\mu$s (f).}
\end{figure}

In figure \ref{Voba} we present the transmissions of both lasers at the corresponding sides of the cell. For the case $I_{pulse}^{0}>I_{CW}^{0}$ (we analyzed range of intensities $I\leq 1$ mW/cm$^{2}$) both lasers are switching each other's resonance signs. The switching happens only at the time when the pulse intensity varies near its maximum, therefore it is a very fast and short optical switch. Results in figures \ref{Voba} (d) and (e) show that both lasers have completely changed the sign of resonance from EIT/EIA to EIA/EIT. Results also show that there is a mirror-symmetry between the transmission of the CW and pulse lasers. Both lasers change the sign of resonance simultaneously. The only exception from this is around the $t=40$ $\mu$s, when both lasers show EIA, when there are also most drastic changes in the evolution of the atomic ensemble.


\begin{figure}[htbp]
\centering\includegraphics[width=0.8 \textwidth]{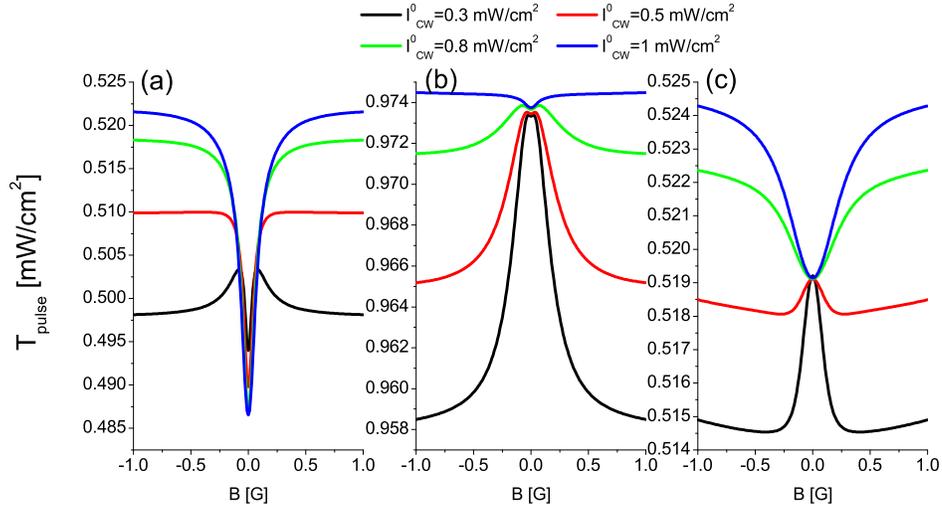}
\caption{\label{Vporedim}
Transmissions of the pulse fields, when the CW laser takes different intensities: $I_{CW}^{0}=0.3$ mW/cm$^{2}$ (black curves), $I_{CW}^{0}=0.5$ mW/cm$^{2}$ (red curves), $I_{CW}^{0}=0.8$ mW/cm$^{2}$ (green curves)  and $I_{CW}^{0}=1$ mW/cm$^{2}$ (blue curves). We observe transmissions in $3$ different moments in time: $t=40$ $\mu$s (a),  $t=50$ $\mu$s (b) and  $t=60$ $\mu$s (c). Maximum intensity of the pulse laser is kept constant $I_{pulse}^{0}=1$ mW/cm$^{2}$.}
\end{figure}

We also calculated transmission of the laser pulse for four values of the CW laser intensities. Results are given in figure \ref{Vporedim} for  $I_{CW}^{0}=0.3$ mW/cm$^{2}$ (black curves), $I_{CW}^{0}=0.5$ mW/cm$^{2}$ (red curves), $I_{CW}^{0}=0.8$ mW/cm$^{2}$ (green curves)  and $I_{CW}^{0}=1$ mW/cm$^{2}$ (blue curves). The intensity of the pulse laser is $I_{pulse}^{0}=1$ mW/cm$^{2}$. We observe pulse's Hanle resonance in transmissions for $3$ different times: $t=40$ $\mu$s (a),  $t=50$ $\mu$s (b) and  $t=60$ $\mu$s (c). Results show that when the intensity of the CW field is varied, the transmission of the pulse laser at the zero magnetic field, for $B=0$, remains constant at all times. The condition for CPT in the Hanle configuration is fulfilled at zero magnetic field, and is, in this case, well established by the pulse's stronger laser. As can be seen in figures \ref{Vporedim} (b) and (c), the rest of the pulse transmission profile, which violates the exact condition for dark-state fulfilled at zero 
magnetic field, can show  either EIT or EIA depending on the strength of the CW field intensity.


\begin{figure}[htbp]
\centering\includegraphics[width=0.8 \textwidth]{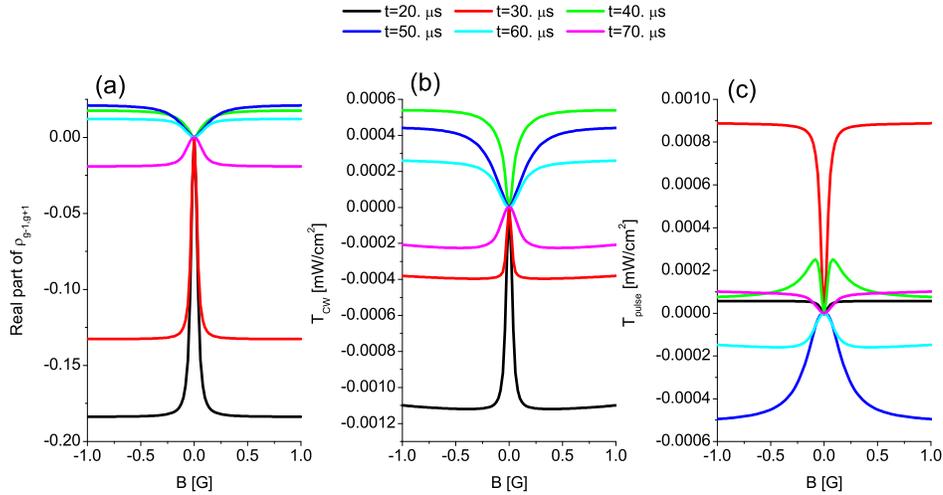}
\caption{\label{GGkohs}
Real part of the ground-state coherences $\rho_{g_{-1,+1}}$ (a), transmissions of the CW laser (b) and pulse laser (c) for the six $6$ times during the time development of the pulse, from $t=20$ $\mu$s to $t=70$ $\mu$s in steps of $10$ $\mu$s. Results for the coherences are calculated at the middle of the cell, $z= 0.05\ m$. The curves in all three graphs are $y$-shifted such that they coincide in $B=0$}
\end{figure}

In figure \ref{GGkohs} we present real part of the ground-state coherences  $\rho_{g_{-1,+1}}$ (a) and compare them with the transmissions of the CW laser (b) and pulse laser (c). Different colors indicate different intensities applied i.e. different moments of the pulse propagation, from $t=20$ $\mu$s to $t=70$ $\mu$s in steps of $10$ $\mu$s, whereas the intensity of the CW laser is constant, $I_{CW}^{0}=0.3$ mW/cm$^{2}$. Curves in all $3$ graphs are shifted along $y$-axis for the easier comparison. We present results at the middle of the cell, since changes along the cell are negligible.  In this configuration, the transmission of the CW laser matches the sign of the ground-state coherences, while of the pulse laser is of the opposite sign. Results presented in figure \ref{GGkohs} confirm that the behavior of the transmissions of both lasers closely follow the behavior of the ground-state coherences.

\section{Conclusion}

We have analyzed the propagation dynamics of two counter-propagating laser fields with Rb atoms in the Hanle configuration, one of which is Gaussian pulse and the other is CW. We demonstrated continuous sign reversal of both lasers in the Hanle configuration. Sign reversal was obtained in two ways, of only CW laser field and of both lasers. The choice lies on the ratio of lasers' intensities. We have also analyzed peculiarities of these two cases, different slopes of pulses, the effect of different intensities and the behavior of the ground-state coherences. The results obtained in this work  may provide a useful reference for further research  and interesting applications of EIT and EIA ultranarrow resonances.

\ack
This work was supported by the Ministry of Education and Science of the Republic of Serbia, under grant number III $45016$.

\end{document}